# Cytoskeletal bundle bending, buckling, and stretching behavior

Mark Bathe[†], Claus Heussinger[†], Mireille M.A.E. Claessens[‡],
Andreas R. Bausch[‡], and Erwin Frey[†]

[†]Arnold Sommerfeld Zentrum für Theoretische Physik and Center for NanoScience
Ludwig–Maximilians–Universität München, Theresienstr. 37, 80333 Munich, Germany
and
[‡]Department of Biophysics E22
Technische Universität München, James Franck Str. 1, 85747 Garching, Germany

**Abstract**

F-actin bundles constitute principal components of a multitude of cytoskeletal processes including stereocilia, filopodia, microvilli, neurosensory bristles, cytoskeletal stress fibers, and the sperm acrosome. The bending, buckling, and stretching behaviors of these processes play key roles in cellular functions ranging from locomotion to mechanotransduction and fertilization. Despite their central importance to cellular function, F-actin bundle mechanics remain poorly understood. Here, we demonstrate that bundle bending stiffness is a state-dependent quantity with three distinct regimes that are mediated by bundle dimensions in addition to crosslink properties. We calculate the complete state-dependence of the bending stiffness and elucidate the mechanical origin of each. A generic set of design parameters delineating the regimes in state-space is derived and used to predict the bending stiffness of a variety of F-actin bundles found in cells. Finally, the broad and direct implications that the isolated state-dependence of F-actin bundle stiffness has on the interpretation of the bending, buckling, and stretching behavior of cytoskeletal bundles is addressed.

**Introduction**

Filamentous actin (F-actin) is a stiff biopolymer that is tightly crosslinked *in vivo* by actin-binding proteins (ABPs) to form stiff bundles that form major constituents of a multitude of slender cytoskeletal processes including stereocilia, filopodia, microvilli, neurosensory bristles, cytoskeletal stress fibers, and the acrosomal process of sperm cells (Fig. 1) (1, 2). The mechanical properties of these cytoskeletal processes play key roles in a broad range of cellular functions—the bending stiffness of stereocilia mediates the mechanochemical transduction of mechanical stimuli such as acoustic waves to detect sound and motion (3, 4), the critical buckling load of filopodia and acrosomal processes determines their ability to withstand compressive mechanical forces generated during cellular locomotion and fertilization (5-7), and the entropic stretching stiffness of cytoskeletal bundles mediates cytoskeletal mechanical resistance to cellular deformation (8). Thus, a detailed understanding of F-actin bundle mechanics is fundamental to gaining a mechanistic understanding of cellular function.

Cells utilize a myriad of ABPs to assemble and crosslink F-actin filaments into bundles of precisely regulated dimensions that range dramatically from several (microvilli, stereocilia, stress fibers) to tens (acrosome, filopodia) and even hundreds (macrochaete neurosensory bristles in *Drosophila*) of microns and from tens (filopodia, microvilli) to hundreds (stereocilia, neurosensory bristles) of constituent filaments (1, 2). Of the multitude of ABPs expressed by the cell, only a small subset is used to crosslink neighboring F-actin filaments in cytoskeletal bundles. Fascin is the predominant ABP in filopodia and neurosensory bristles, plastin is prevalent in microvilli and stereocilia, scruin in the *limulus* sperm acrosome, and $\alpha$–actinin predominates in cytoskeletal stress fibers. Each ABP has a distinct mechanical shear stiffness that has been demonstrated to strongly affect F-actin bundle bending stiffness *in vitro* (9) and *in vivo*: stereocilia predominant in plastin exhibit weak, *decoupled* bending (4, 10), $\kappa_B \sim n\kappa_f$, in which the $n$ constituent fibers with bending stiffness $\kappa_f$ bend independently like the sheets in a loose stack of paper, whereas the *limulus* sperm acrosome prevalent in scruin exhibits a much higher *fully coupled* bending stiffness (11), $\kappa_B \sim n^2\kappa_f$, like a homogeneous mechanical beam. It is not obvious *a priori*, however, whether these drastically different regimes of bending stiffness are determined only by ABP *type*, or whether bundle dimensions play a commensurate leading role. Moreover, an additional bending regime that is intermediate to decoupled and fully coupled bending has been observed in F-actin (9), microtubule (12, 13), and carbon nanotube (14) bundles, however its nature and mechanical origin remain obscure.

In this article, we demonstrate that F-actin bundles have three distinct bending regimes that are mediated by both *ABP type* and equally importantly by *bundle dimensions*—namely diameter and length. We isolate the origin of the third, *intermediate* regime to decoupled and fully coupled bending and demonstrate that it interestingly exhibits a bending stiffness that, unlike the other regimes, is proportional to crosslink shear stiffness and bundle length. We derive a generic set of design parameters that delineates the three bending regimes and use it to make novel predictions for the bending behavior of cytoskeletal bundles that are not easily amenable to experimental measurement. Finally, the direct and broad implications that these results have on the interpretation of the bending, buckling, and stretching behavior of the multitude of cytoskeletal bundles found in cells are addressed.

## Model

We consider generic fiber bundles of length $L$ that consist of $n$ cubically- or hexagonally-packed fibers, as is typical of highly crosslinked F-actin, MTs, and SWNTs (10, 15, 16) (Fig. 2A). Each cylindrical fiber is characterized geometrically at a coarse-grained molecular-scale by its diameter, $d_f$ [m$^2$], and contour length, $L_f$ [m]. Fibers run the full length of the bundle $(L_f = L)$ and are modeled mechanically as extensible Euler–Bernoulli beams (or extensible worm-like polymers) with stretching stiffness, $k_f := E_f A_f / L_f$ [N/m], and isotropic transverse bending stiffness, $\kappa_f := E_f I_f$ [Nm$^2$]. $E_f$ [N/m$^2$] is the effective Young's modulus of the fiber, $A_f$ is its cross-sectional area, and $I_f$ [m$^4$] is the moment of inertia of its cross-sectional area with respect to its neutral axis.[§**] Fibers are irreversibly crosslinked to their nearest-neighbors by discrete inextensible crosslinks with shear stiffness, $k_\parallel$ [N/m], length, $t$ [m], and axial spacing, $\delta$ [m]. Crosslinks therefore constrain transverse fiber deflections to be equal but allow interfiber relative slip. The consideration of ordered fiber bundles simplifies our analyses to in-plane bending of $N = \sqrt{n}$ fiber *layers* that are crosslinked to their nearest neighbors in- and out-of-plane (Fig. 2A), where the corresponding 3D bundle bending stiffness is related simply to its 2D counterpart by, $\kappa_B := N \kappa_{B(2D)}$.[††] Various types of biological fiber bundles have been modeled previously along similar lines (4, 10, 12, 17).

---

[§] For molecular-scale objects, $k_f$ and $\kappa_f$ are fundamental independent observables that may be measured experimentally, whereas, $E_f$, $A_f$, and $I_f$ are continuum mechanics equivalents that are ill-defined at the molecular-scale and thus only *effective* in their nature.

[**] The neutral *surface* of a beam is the surface on which the bending-induced axial strain is zero. The intersection of that surface with any beam cross-section defines the neutral *axis* of the beam.

[††] Effects of out-of-plane shear deformations present in hexagonally-packed bundles during planar bending, as well as finite-size geometric boundary effects, are ignored to leading order.

Bundle deformations are characterized solely by $r_\perp(x)$, the transverse deflection of the bundle neutral surface at axial position $x$ along its backbone. In contrast, fiber deformations are characterized by the linear superposition of this transverse deflection and their mean axial extension, $\bar{u}^{(k)}(x) = \frac{1}{A_f} \int_{A_f} u^{(k)}(x, y) dA$, where $u^{(k)}(x, y)$ is the axial displacement field in the $k^{th}$ fiber $(k = 1, 2, ..., N)$ and the overbar denotes fiber-cross-sectional-average (Fig. 2B). The associated axial strain field, $\varepsilon^{(k)} := u_{,x}^{(k)}$, in the $k^{th}$ fiber similarly consists of a linear superposition of bending- and stretching-induced contributions, $\varepsilon^{(k)} = -r_{\perp,xx} \tilde{y} + \bar{u}_{,x}^{(k)}$, where $\tilde{y}$ denotes distance from the *fiber* neutral axis, a subscript comma is used to denote differentiation, and the standard small-displacement approximation $\rho \approx (r_{\perp,xx})^{-1}$ has been used for the neutral surface radius of curvature, $\rho$ (18).

Crosslink shear displacements, $\nu$, result from both *stretching* as well as *plane-cross-section-rotations* of neighboring fibers, $\nu_j^{(k)} = \bar{u}^{(k)}(x_j) - \bar{u}^{(k-1)}(x_j) + d_f r_{\perp,x}$, $(k = 2, 3, ..., N)$, where $j$ labels the crosslink at axial position $x_j = j\delta$ $(j = 1, 2, ..., L/\delta)$ and $(t \ll d_f)$ has been assumed. The shear displacement may be written equivalently in terms of fiber mean axial strain and inverse radius of curvature,

$$\nu_j^{(k)} = \int_0^{x_j} (\bar{\varepsilon}^{(k)} - \bar{\varepsilon}^{(k-1)} + d_f r_{\perp,xx}) dx.$$

While the stretching and bending stiffness of F-actin (19-21), MTs (13, 20), and SWNTs (22, 23) are experimentally known, the shear stiffness of fiber crosslinks is often unknown. One exception is provided by the recent measurements of Claessens et al., (9), in which an effective $k_\parallel$ was measured for the ABPs plastin, fascin, and $\alpha$–actinin. In other cases, $k_\parallel$ may in principle be calculated directly using atomistic-based simulation methods or measured using micromanipulation techniques. The spacing between fibers, $t$, can be measured from crystal structures (15, 24, 25) and $\delta$ can be determined from chemical equilibrium and fiber packing considerations (9, 16).

Some biological crosslinks such as the ABPs fascin and plastin have finite off-rates, $k_{off} \sim 0.1 - 1 \ \text{s}^{-1}$ (26), and are therefore only irreversibly-bound on loading or deformation time scales that are shorter than $k_{off}^{-1}$. On longer time scales crosslinks may dissociate and rebind, thereby relaxing their shear deformation energy, such as in the coiled packing of the F-actin bundle of the sperm acrosome in which kinking via crosslink unbinding and subsequent inter-filament slip occurs (27). While the effects of crosslink unbinding/rebinding are of interest for some biological loading scenarios, they are beyond the scope of the present work.

In addition to finite off-rates, real crosslinks also have a finite extensibility $k_\perp$ [N/m] that could in principle allow for fiber bending undulations. Typical crosslinking proteins have an extensional stiffness, $k_\perp \sim 1 \ \text{N/m}$ (28), however, that restricts the wavelength of fiber undulations to lengths at or below the typical crosslink distance, $\delta$.[‡‡] Because we consider bundles for which, $\delta \ll l_p$, where $l_p$ is the bare persistence length of the fiber, the associated fluctuation amplitude, $r_\perp \sim \delta^{3/2} / l_p^{1/2}$ (29), may safely be neglected. Thus, fibers typically remain tightly packed and ordered, as demonstrated by electron microscopy (16), and the assumption of inextensible crosslinks is justified.

---

[‡‡] Crosslinks suppress fiber bending undulations to wavelengths, $\lambda \leq \lambda_{max} := (\kappa_f \delta / k_\perp)^{1/4}$, where $\lambda_{max} \approx 10 \ \text{nm}$ for F-actin with $\kappa_f \approx 7 \times 10^{-26} \ \text{Nm}^2$ (19, 20). The minimum axial distance between co-planar crosslinks in hexagonally-packed F-actin bundles is 37.5 nm (16). The associated transverse fluctuation of F-actin is $r_\perp \sim 1 \ \text{nm}$, which is much less than the inter-axial spacing between fibers, $(d_f + t) \geq 10 \ \text{nm}$ (16).

The three-dimensional bundle bending stiffness can in general be expressed as a function of all the independent parameters in the model, $\kappa_B(n, L, k_f, \kappa_f, k_\parallel, \delta, t)$, which in dimensionless form may be written, $\kappa_B^* = \kappa_B^*(n, k_\parallel L^3 / \kappa_f, k_f L^3 / \kappa_f, L / \delta)$, where $\kappa_B^* := \kappa_B / \kappa_f$ and the limit of short crosslinks $(t \ll d_f)$ has been applied. We will shortly demonstrate, however, that $\kappa_B^*$ depends only on the two independent parameters, $n$ and the fiber-coupling parameter,

$$\alpha := \frac{k_\parallel L}{k_f \delta} \qquad (1)$$

The fiber coupling parameter is evidently a measure of the competition between *crosslink shearing* and *fiber stretching*, where $L / \delta$ is the number of crosslinks per fiber.

### Numerical analysis

To elucidate the detailed mechanics of fiber-bundle bending, we begin by examining the bending response of model fiber bundles subject to simple three-point bending computationally using the finite element (FE) method (Materials and Methods).[§§] In analogy with experiment, $\kappa_B$ is evaluated as a function of increasing fiber number $n$, for bundles of fixed $\alpha$, which is akin to fixing the fiber and crosslink properties (Fig. 3A). Decoupled bending characterized by linear scaling is observed for small $\alpha$ and fully coupled bending for large $\alpha$. Interestingly, between these two limits we also observe an *intermediate* range of $\alpha$ that displays a smooth crossover from quadratic-to linear-scaling in $n$. This is in contrast to a bending stiffness that is characterized by an $\alpha$-dependent exponent $a$, $\kappa_B \sim n^a \kappa_f$ [$1 < a(\alpha) < 2$] (4, 5). This crossover was recently observed experimentally in F-actin bundles crosslinked by fascin using a controlled variation of $n$ at fixed fascin concentration (9). Re-plotting $\kappa_B^*$ as a function of $\alpha$ indicates that this range is in fact part of a distinct *intermediate regime* where $\kappa_B^*$ increases with increasing $\alpha$ (Fig. 3B). Moreover, any bundle that exhibits fully coupled bending behavior at any given $\alpha$ necessarily transitions into this regime with increasing bundle size, $n$. In what follows we perform a scaling analysis that considers the energetic competition between *fiber stretching* and *crosslink shearing* to elucidate the physical origin of the crossovers between each regime and to delineate their boundaries in $(n, \alpha) - \text{space}$.

---

[§§] Three-point beam bending refers to pinning or clamping a beam at its ends and applying a transverse point load at its center. The resultant load-deflection yields a measure of its apparent bending stiffness (Materials and Methods).

### Scaling analysis

Consider a generic fiber bundle with a fixed characteristic radius of curvature, $\rho \approx (r_{\perp,xx})^{-1}$. In the decoupled limit individual fibers bend equally without stretching, whereas in the fully coupled limit crosslinks resist shear deformation so that fibers are forced to stretch and compress in addition to bend (Fig. 2B). Differences in fiber deformations in the decoupled, fully coupled, and intermediate regimes are thus manifest at fixed radius of curvature solely in differences in mean *fiber stretching*.

Accordingly, to isolate the crossover from the fully coupled to the intermediate regime we impose an infinitesimal stretching deformation, $\delta\bar{\varepsilon}^{(k)}$, that relaxes extensionally the fibers and thereby reduces the total fiber stretching energy, $W_s$, at the expense of an increase in crosslink shearing energy, $W_\parallel$. $\delta\overline{\varepsilon}^{(k)}$ is a characteristic deformation that is constant along the bundle axis but may differ between fiber layers, $k$. The crossover between the fully coupled and intermediate regimes is then determined by the point at which crosslink shearing becomes favorable to fiber stretching,

$\delta W_s[\delta\overline{\varepsilon}^{(k)}] = \delta W_\parallel[\delta\overline{\varepsilon}^{(k)}]$, where $\delta W_s[\delta\overline{\varepsilon}^{(k)}] = N \sum_{k=1}^{N} \int_0^L dx \overline{F}^{(k)} \delta\overline{\varepsilon}^{(k)}$ is the variation in stretching energy and $\delta W_\parallel[\delta\overline{\varepsilon}^{(k)}] = N \sum_{k=2}^{N} \sum_{j=1}^{L/\delta} F_{\parallel j}^{(k)} \delta v_j^{(k)}$ is the variation in crosslink shearing energy associated with the imposed relaxation $\delta\overline{\varepsilon}^{(k)}$ that results in the crosslink displacement, $\delta v_j^{(k)} = (\delta\overline{\varepsilon}^{(k)} - \delta\overline{\varepsilon}^{(k-1)}) x_j$. The calculation of these energy variations requires that the mean-fiber-stretching-, $\overline{F}^{(k)}$, and crosslink-force, $F_{\parallel j}^{(k)}$, conjugate to the deformations $\delta\overline{\varepsilon}^{(k)}$ and $\delta v_j^{(k)}$ be evaluated, which we turn to next.

The mean axial force in the $k^{th}$ fiber is related via Hooke's law to its mean axial strain, $\overline{F}^{(k)} = E_f A_f \overline{\varepsilon}^{(k)}$, which in the fully coupled regime increases linearly with distance, $y$, from the *bundle* neutral axis, $\overline{\varepsilon}^{(k)} = -y(k) r_{\perp,xx} = -(k - \frac{N}{2} - \frac{1}{2}) d_f r_{\perp,xx}$, so that, $\overline{F}^{(k)} = -E_f A_f (k - \frac{N}{2} - \frac{1}{2}) d_f r_{\perp,xx}$, like in a homogeneous Euler–Bernoulli beam (Fig. 2B). The limit of small crosslinks $(t \ll d_f)$ and $(N = even)$ have been assumed here for simplicity without loss of generality. It is precisely this fiber stretching force that gives rise to the additional bundle bending moment and higher associated bundle bending stiffness in the fully coupled regime. The crosslink force, $F_{\parallel j}^{(k)}$, is linearly related to its shear displacement via, $F_{\parallel j}^{(k)} = k_\parallel v_j^{(k)}$, which is given by, $v_j^{(k)} \sim d_f r_{\perp,xx} x_j$, so that, $F_{\parallel j}^{(k)} \sim k_\parallel d_f r_{\perp,xx} x_j$, where a constant characteristic radius of curvature has been assumed in evaluating $\upsilon$, consistent with the present scaling picture. Note the differences between the expressions for the fiber axial force and the crosslink shear force: The former increases through the bundle *cross-section* whereas the latter increases along the bundle *axis*.

Variations in fiber stretching and crosslink shearing energy associated with the imposed relaxation $\delta\overline{\varepsilon}^{(k)}$ may now be calculated using the above results to yield,

$$\delta W_s \sim N A_f E_f d_f r_{\perp,xx} L \sum_{k=1}^{N} k \delta\overline{\varepsilon}^{(k)} \text{ and } \delta W_{\parallel} \sim N k_{\parallel} d_f r_{\perp,xx} \sum_{k=2}^{N} (\delta\overline{\varepsilon}^{(k)} - \delta\overline{\varepsilon}^{(k-1)}) \sum_{j=1}^{L/\delta} x_j^2 ,$$

which may be re-written, $\delta W_{\parallel} \sim N k_{\parallel} d_f r_{\perp,xx} (L^3 / \delta) \sum_{k=2}^{N} (\delta\overline{\varepsilon}^{(k)} - \delta\overline{\varepsilon}^{(k-1)})$, after evaluation of the summation over crosslinks. Equating the resultant increase in crosslink shear energy with the decrease in fiber stretching energy and imposing arbitrary $\delta\overline{\varepsilon}^{(k)}$ determines the location of the crossover, $N^2 E_f A_f \sim k_{\parallel} L^2 / \delta$, which may be re-written, $\alpha \sim n$. Thus, the crossover from the fully coupled regime to the intermediate regime occurs at higher $\alpha$ for larger bundles, *n*. This result is due to the fact that in the fully coupled regime the fiber stretching energy scales with bundle *diameter* whereas the crosslink shearing energy scales with bundle *length*.

A similar analysis applies to the decoupled limit except that fibers are initially unstressed axially in the ground state. Finite element results indicate that axial stretching is first induced in fibers at the outer boundary of the bundle in order to minimize the associated increase in $\delta W_s$, because inner fibers then remain in their relaxed state. This leads directly to a crossover that is bundle-diameter- and thus *n*-independent, which is given by the condition, $\alpha \sim 1$. Comparison of the crossovers between the decoupled–intermediate ($\alpha \sim 1$) and fully-coupled–intermediate ($\alpha \sim n$) regimes computed with the finite element model confirms the validity of the foregoing scaling arguments (Fig. 3B and Fig. 3B Inset), with some deviations for small $n$. Introduction of the finite-size, $t$, of the crosslinks increases the absolute value of the fully coupled bending stiffness but it does not affect this scaling behavior.

### *Analytical solution*

The fiber bundle model admits an analytical solution to $\kappa_B$ employing a continuum energetic approach. The total elastic energy of the fiber bundle, $W[r_\perp(x), u^{(k)}(x)]$, is decomposed into fiber bending, $W_b$, fiber stretching, $W_s$, and crosslink shearing, $W_\parallel$, contributions. The bending contribution is given by a linear superposition of the standard worm-like chain bending energy for each independent fiber, $W_b = \frac{1}{2} n \kappa_f \int_0^L r_{\perp,xx}^2 dx$, because transverse fiber-displacements are equal. The fiber stretching energy is given by the axial strain energy, $W_s = \frac{1}{2} N E_f A_f \sum_{k=1}^{N} \int_0^L \left(\overline{u}_{,x}^{(k)}\right)^2 dx$. Finally, crosslink shear energy is associated with crosslink deformation that results from neighboring fiber bending and stretching, $W_\parallel = \frac{1}{2\delta} N k_\parallel \sum_{k=2}^{N} \int_0^L \left(v^{(k)}(x)\right)^2 dx$.

The theoretical model contains $N$ internal stretching degrees of freedom $\overline{u}^{(k)}$ in addition to the transverse bundle deflection, $r_\perp$, which is the principal macroscopic observable of interest in measuring the bundle response. Accordingly, the bundle energy is minimized with respect to $\overline{u}^{(k)}$ to arrive at an *effective* bundle bending energy that depends only on $r_\perp$, from which the bundle bending stiffness may be obtained. The minimization is performed using a Fourier decomposition of the functions $r_\perp(x) = \sum_j r_j \sin(q_j x)$ and $\bar{u}^{(k)}(x) = \sum_j u_j^{(k)} \cos(q_j x)$ with associated wavenumbers

$q_j = j\pi / L$ $(j = 1, 2, ...)$ (Supplementary Material)(30). The resulting *mode-number-dependent stiffness* is,

$$\kappa_B(n, \alpha, q_j) = \kappa_f n \left( 1 + \frac{n-1}{1 + c(q_j)\dfrac{n + \sqrt{n}}{\alpha}} \right) \tag{2}$$

which has been derived previously for the special case of two filaments in the context of

DNA mechanics (17). It depends on $q_j$ through the non-dimensional factor

$c(q) = (qL)^2 / 12$ and on the design parameters *n* and $\alpha$ isolated previously using scaling

analysis. In three-point bending at zero temperature the bending stiffness is well-

approximated by Eq. (2) without the mode-number dependence and a constant factor

$c = 1$ for pinned ends and $c = 4$ for clamped ends, in quantitative agreement with the

Finite Element results. This expression reduces to the decoupled and fully coupled bending stiffness in the limits $(\alpha \ll 1)$ and $(\alpha \gg n)$, respectively, and exhibits the scaling, $\kappa_B \propto nA_f L_f^2 k_\parallel / \delta$, in the intermediate regime, $(1 \ll \alpha \ll n)$, which is *independent*

of the mechanical properties of the underlying fibers. This demonstrates that the

intermediate regime is dominated by *shear-deformation* of the crosslinks, so that

*intermediate* and *shear-dominated* may be used interchangeably. This is in contrast to the decoupled and fully coupled regimes, in which the crosslink shear stiffness is effectively equal to zero and infinity, respectively.

The *q*-dependence of $\kappa_B$ demonstrates that it is an *apparent* material property that depends on the nature in which the bundle is probed. This is in contrast to a standard worm-like polymer, which is defined as having an *intrinsic* bending stiffness that is mode-number-independent (31). Thus, inference of $\kappa_B$ from "macroscopic" bundle observables such as the mean-square end-to-end distance, the zero-temperature force-deflection relation, or the fluctuation spectrum by associating the bundle with an equivalent worm-like polymer will yield different apparent values for $\kappa_B$ (30).

### Connection to Timoshenko theory

Fiber bundles consisting of SWNTs (14, 32) and MT protofilaments (13, 33) have recently been analyzed using Timoshenko beam theory[***]. In this approach, the heterogeneous microstructure of the bundle is ignored so that the bundle can instead be treated as a single homogeneous medium with effective macroscopic geometric and mechanical properties. The bundle stiffness computed from Timoshenko theory for three-point bending with pinned boundary conditions may be written (34),

$\kappa_B = E_B I_B \left[ 1 + 12 E_B I_B / \beta G_B A_B L_B^2 \right]^{-1}$, where $G_B$ is the effective bundle shear modulus and $\beta$ is a cross-section-dependent shear-correction factor. To make a connection with the microscopic fiber-bundle theory employed in this work, the interlayer crosslink shear force is assumed to be constant across the bundle cross-section and equal to the macroscopic effective shear stress, $\tau^{macro} = G_B \gamma_B := \tau^{micro} = k_\parallel v / d_f \delta$, where $\gamma_B$ is the equivalent macroscopic bundle shear strain. The microscopic interlayer slip is also assumed to be transversely invariant and related to the macroscopic shear strain via, $\gamma_B = \nu / d_f$. Substitution yields, $\kappa_B = n^2 \kappa_f \left(1 + n/\alpha\right)^{-1}$, which is identical to the fiber-based model result when the limit $(n \gg 1; \ \alpha \gg 1)$ is applied.

---

[***]Microtubules have been analogized to "bundles" by considering protofilaments as the fibers and inter-protofilament bonds as effective crosslinks.

Thus, Timoshenko theory converges to the same fully coupled bundle bending stiffness as the microscopic-based theory when, $(\alpha \gg n)$, and crosses-over to the shear-dominated regime when, $(\alpha \sim n)$ (Fig. 3B). Unlike the microscopic theory, however, Timoshenko theory does not asymptote to the decoupled bending regime when, $(\alpha \ll 1)$, and it is only asymptotically correct for large bundles, $(n \gg 1)$, because it does not explicitly account for the heterogeneous underlying fiber structure of the bundle (Fig. 3B).

### *Application to F-actin bundles*

The bending stiffness of F-actin bundled by fascin, plastin, $\alpha$–actinin, or non-specific PEG-induced depletion forces was recently measured experimentally using an *in vitro* droplet assay in which F-actin bundles form compact stable rings (9). In that work, the dependence of bundle stiffness on bundle diameter $n$ was systematically explored for several ABP concentrations. Here, we focus on fascin and instead explore the effects of bundle length and fascin concentration on $\kappa_B$ for a single bundle diameter, $n = 27 \pm 3$ (Fig. 3C). Uncorrelated variation of fascin concentration, which varies the crosslinker spacing $(40 \text{ nm} \leq \delta \leq 400 \text{ nm})$, and bundle length $(24 \mu\text{m} \leq L \leq 55 \mu\text{m})$ results in an increase in bundle stiffness from decoupled (at small $L^2 / \delta$) to intermediate regime and finally fully coupled (at large $L^2 / \delta$) bending. Fitting Eq. (2) to the data using $c = 4$ and substituting the stretching stiffness of F-actin $(k_f L_f = 4.4 \times 10^{-8} \text{ N})$ (21) yields a unique value for the effective shear stiffness of fascin, $k_\parallel \approx 10^{-5}$ N/m, without any adjustable parameters.

## Bending stiffness state diagram

The identification of the generic design parameters $(n, \alpha)$ allows for the bending stiffness of a broad range of biological and synthetic fiber bundles to be placed on a universal bending stiffness state-diagram for F-actin-, MT-, and SWNT-based bundles (Fig. 4). Maximal bundle bending stiffness is achieved by ensuring fully coupled bending $(\alpha \gg n)$, whereas maximal bundle compliance is achieved by decoupled bending $(\alpha \ll 1)$. In the shear-dominated regime $(1 \ll \alpha \ll n)$ crosslink concentration or bundle length may be varied to tune $\kappa_B$ by orders of magnitude.

The sperm acrosomal process functions to mechanically penetrate the outer jelly coat of the egg cell during fertilization (7, 11). The *limulus* (horseshoe crab) sperm acrosome consists of a tapered bundle of 15–80 hexagonally-packed F-actin fibers that are tightly crosslinked by scruin and run the full length ($L \approx 50 \ \mu$m) of the bundle. Macroscopic measurements of its bending stiffness have been made using hydrodynamic flow (11), where it was determined that the bundle exhibits fully coupled bending. This independent macroscopic observation is consistent with the *a priori* prediction of the fiber-based model, in which the ranges in $\alpha$ and $n$ are determined from the parameters probed experimentally (Fig. 4). The shear stiffness of fascin is used as an estimate for scruin, although the molecular structure and interfilament packing of the latter suggest that it is considerably stiffer (15).

Vertebrate hair cell stereocilia are finger-like projections in the inner ear that serve as mechanochemical transducers for sound and motion (Fig. 1A). Ranging from 1–10 $\mu$m in length, each stereocilium consists of up to 900 hexagonally-packed F-actin filaments crosslinked predominantly by plastin (1, 2, 35). Macroscopic measurements of the bending stiffness of hair cell stereocilia bundles and of the root of individual stereocilia made using microneedle manipulation (4) yielded decoupled bending behavior. Together with their short length, the low stiffness of plastin, $k_\parallel \leq 10^{-6}$ N/m (9), places their theoretical stiffness deep in the decoupled regime, consistent with these experimental observations (Fig. 4).

Knowledge of the microstructure and the filament and crosslink mechanical properties may be used to also make novel predictions of $\kappa_B$ for cytoskeletal processes that have not been measured experimentally. Of course, *in vivo* F-actin bundles are typically crosslinked by more than one ABP type (36), however one ABP is prevalent in each process and is therefore expected to dominate the bundle response (1, 2).

Brush-border microvilli ($n \approx 20-30$; $L \approx 1-5$ $\mu$m) are passive cellular processes that predominate in plastin and serve primarily to increase the apical surface area of intestinal epithelial cells (1, 2) (Fig. 1D). Cytoskeletal stress fibers ($n \approx 10-40$; $L \approx 1-10$ $\mu$m) predominate in $\alpha$–actinin $k_\parallel \approx 10^{-5}$ N/m (9) and function mechanically to enhance the tensile stiffness of cells. Each of these processes is predicted to exhibit decoupled bending due to its relatively short length. Filopodia are active F-actin bundles present at the leading edge of motile cells and neuronal growth cones that increase in length during locomotion and growth (2) (Fig. 1B). Consisting of 10–30 filaments, they are predominantly crosslinked by fascin and typically range from 1–10 $\mu$m, but may reach lengths of up to 30–40 $\mu$m in certain cases such as in the sea urchin embryo (5, 37). As a final F-actin bundle example, we consider the 11 fascin-crosslinked bundles constituting the *Drosophila* neurosensory bristle. Each bundle is ≈400 microns long and contains 500–700 filaments in macrochaetes (38, 39) (Fig. 1C). Using their full length, these bundles are predicted to lie at the interface of the fully coupled and intermediate regimes, despite their large diameter. Early in development, however, bristles consist of short overlapping bundle modules $(L_f \approx 3\ \mu\mathrm{m})$ (38). At this early stage the fiber fracture density, $\phi := L / L_f \approx 100$, is less than the *critical fracture density*, $\phi^* \sim n \sim 10^2 - 10^3$, below which we find the fully-coupled–intermediate regime transition to be unaffected by fracture (Supplementary Material). This critical value has its origin in the fact that nearest-neighbor fibers can carry the pre-existing axial load of a fractured fiber. Direct bending stiffness measurements would be of interest to verify this interpretation.

Finally, noting that the bundle model employed in this work is completely generic, we also include in the state diagram MT bundles from outer pillar hair cells for which the interlayer shear modulus has been measured using micromanipulation and a fiber-based model $(n \approx 1000 - 3000;\ L \approx 70 - 120\ \mu\mathrm{m};\ k_\parallel / \delta \approx 1\ \mathrm{kPa})$ (Fig. 1F) (12), and uncrosslinked $(n \approx 10 - 200;\ L \approx 0.1 - 0.4\ \mu\mathrm{m};\ k_\parallel / \delta \approx 1\ \mathrm{GPa})$ (Fig. 1G) (14) and irradiation-crosslinked $(n \approx 10 - 200;\ L \approx 0.1 - 0.4\ \mu\mathrm{m};\ k_\parallel / \delta \approx 200\ \mathrm{GPa})$ (32) SWNTs that were probed using AFM and analyzed using macroscopic Timoshenko theory to determine the apparent macroscopic bundle shear modulus, $G_B \approx k_\parallel / \delta$.

**Implications for in situ mechanical function**

Filopodia and the sperm acrosome are amongst two F-actin-based cytoskeletal processes that are subject to potentially high compressive forces *in vivo* during locomotion, growth, and fertilization (5, 6, 40). Bundles subject to axial compression *in situ* will lead to structural failure at a critical load that is determined by the Euler buckling limit, $F_{crit} = \pi^2 \kappa_B / 4L^2$. Interestingly, whereas $F_{crit} \propto 1/L^2$ in the decoupled and fully coupled regimes, $F_{crit}$ becomes $L$-independent in the shear-dominated regime, because $\kappa_B \propto L^2$ there. Thus, dynamic cytoskeletal fiber bundle processes such as filopodia may grow in length without decreasing their critical buckling load in the intermediate regime, and thereby become length-limited only once they reach the fully coupled regime, when $F_{crit}$ becomes strongly dependent on length. This interesting mechanical feature of fiber bundles is completely generic, of course, and may well be exploited in the design of novel materials.

The entropic stretching stiffness of F-actin bundles is thought to play an important role in determining the elasticity of crosslinked F-actin-ABP networks (8). In the decoupled and fully coupled bending regimes the entropic stretching stiffness of a fiber bundle is similar to that of a semi-flexible wormlike chain, $k_e \sim \kappa_B^2 / k_B T L^4$, where $k_B T$ [J] is thermal energy. Accordingly, its dependence on $n$ in the decoupled regime, $k_e \sim n^2 \kappa_f / k_B T L^4$, is drastically different than that in the fully coupled regime, $k_e \sim n^4 \kappa_f^2 / k_B T L^4$, which has direct consequences for the plateau modulus of F-actin networks. Together, these examples demonstrate that consideration of the *state-* *dependence* of $\kappa_B(n,\alpha)$, as elucidated in this work, are crucial for the proper interpretation of *in situ* bundle mechanical function.

ABPs are complex, hierarchically structured macromolecules that may dissociate and rebind as well as exhibit highly nonlinear force–extension response depending on the time- and length-scales probed (41). Accordingly, the coupling parameter $\alpha$ is in fact a nonlinear function that depends on the *degree* and *time-scale* of crosslink deformation. Thus, bundles in one bending regime may potentially switch to other regimes depending on the deformations imposed *in situ*, and this may be an important modeling consideration. Investigation of the nonlinear- and time-dependent-response of fiber bundles provides a rich avenue of investigation that will require careful and controlled experimentation together with atomistic modeling to unravel in the future.

## Conclusion

Crosslinked F-actin bundles are key structural components of a broad range of cytoskeletal processes. To date, a common conception has been that these bundles display two limiting bending behaviors that depend solely on the stiffness of intervening crosslinks: decoupled or fully coupled. Here, we demonstrate that their bending behavior is considerably more intricate. Their bending regime can be switched by varying global bundle dimensions, namely diameter or length, the shear stiffness of intervening crosslinks, as well as the stretching stiffness and length of constituent fibers. We isolate the design parameters $n$ and $\alpha$ that characterize the bending regime of generic fiber bundles and use them to recast the stiffness of a broad range of cytoskeletal bundles on a universal bending stiffness state diagram, making novel predictions for cellular processes that are not easily amenable to experimental measurement. Experimental bending stiffness of fascin-crosslinked F-actin bundles validates our interpretation of F-actin bundle mechanics, which has important implications for the bending, buckling, and stretching behavior of numerous cytoskeletal processes. Our results are completely generic in nature and thus are equally applicable to bundles of microtubules or carbon nanotubes as they are to F-actin.

## Materials and Methods

**Finite element modeling.** Fibers are discretized identically in 2D using 2-node Hermitian beam elements with nodal degrees of freedom, $\{u_i, w_i, \theta_i\}$, where $u_i$ is axial displacement, $w_i$ is transverse deflection, and $\theta_i$ is in-plane rotation (42). Nodes on adjacent fibers are constrained to have equal transverse deflection. Crosslink shear stiffness is modeled using a general 2-node finite element that couples beam element nodes on nearest-neighbor fibers with stiffness matrix, $K_{ij} = \partial^2 E / \partial x_i \partial x_j$, where $k$ denotes fiber number and $x_i$ denotes the nodal degree of freedom $\{x_i : u^{(k)}, u^{(k-1)}, \theta^{(k)}, \theta^{(k-1)}\}$. The crosslink shear energy function is, $E = (k_\parallel / 2) \left[ (u^{(k)} - u^{(k-1)}) + (d_f / 2)(\theta^{(k)} + \theta^{(k-1)}) \right]^2$, where $k_\parallel$ is normalized properly to account for discretization. Three-point bending is simulated by applying pinned or clamped boundary conditions to the bundle ends and applying a transverse unit point load at the bundle mid-point, yielding the apparent worm-like chain bending stiffness, $\kappa_B := P L^3 / a w_{L/2}$, where $a = 48$ and $a = 192$ for pinned and clamped ends, respectively. Simulations are performed using the commercial Finite Element Software ADINA (ver. 8.2.0).

## Acknowledgements

We thank M. Rusp for the actin preparation, M. Schlierf for his help with the protein expression and purification, D. Vignjevic for the kind gift of recombinant fascin plasmids, T. Svitkina for providing a high resolution version of Fig. 1b, and Tobias Munk for his critical reading of the manuscript. Funding from the DFG in the form of Sonderforschungsbereich 413 and from the Alexander von Humboldt Foundation in the form of a post-graduate research fellowship (to MB) is gratefully acknowledged.

**Figure Legends**

**Fig. 1.** Fiber bundles consisting of F-actin: (a) Ciliary bundle from the sensory epithelium of a bullfrog saccule consisting of about 60 stereocilia. Courtesy David P. Corey and John A. Assad. (b) Filopodium protruding from the lamellipodium of a mouse melanoma cell. Reproduced from (43) by copyright permission of The Rockefeller University Press. (c) Drosophila neurosensory micro- and macrochaete bristles. Reproduced from (44) with the permission of The American Society for Cell Biology. (d) Epithelial microvilli. Microtubule fiber bundles: (e) Demembranated flagellum of Chlamydomonas flagellar axoneme. Courtesy Harold J. Hoops and George B. Witman, unpublished. (f) Cross-sectional view of an outer pillar hair cell bundle from the mammalian inner ear (12). A carbon nanotube fiber bundle: (g) SWNTs bundled noncovalently by van der Waals interactions. Reprinted with permission from (14). Copyright (1999) by the American Physical Society.

**Fig. 2.** Theoretical bundle model (not drawn to scale). **A)** Crosslinked fiber bundle with $n = 16$ fibers. Discrete crosslinks (blue) couple nearest-neighbor fibers mechanically in stretching and bending. **B)** (left) Deformed backbone of a fiber bundle subject to in-plane bending; (middle) close-up view of three typical fibers showing fiber and crosslink deformations in (faded gray lines) decoupled and (solid black lines) fully coupled bending; (right) transverse distributions of fiber axial displacement, $u^{(k)}(x, y)$, and strain, $\varepsilon^{(k)}(x, y)$, fields and (arrows) the mean axial displacement, $\bar{u}^{(k)}(x)$, in (faded gray lines) decoupled and (solid black lines) fully coupled bending. See text for details.

**Fig. 3.** Bundle bending stiffness. **A)** Theoretical $\kappa_B^* := \kappa_B / \kappa_f$ dependence on $n$ for constant $\alpha = \{10^{-1}, 10^0, 10^1, 10^2, 10^3, 10^4\}$ (bottom to top). Thick dotted red lines denote (bottom) decoupled and (top) fully coupled bending regimes. **B)** $n$-normalized dependence of theoretical $\kappa_B^*$ on $\alpha$ at constant $n = \{4, 9, 16,..., 100\}$ (bottom to top). Dotted lines are calculated from Timoshenko theory for (bottom) $n = 4$ and (top) $n = 100$ fibers. **C)** Bending stiffness of fascin-crosslinked F-actin bundles $(n = 27 \pm 3)$. Experimental bundle stiffness (symbols) is measured using a microemulsion droplet system according to (9) for a range of fascin concentrations with corresponding mean spacings, $\delta$: (black circles) 40 nm, (blue squares) 56 nm, (red diamonds) 68 nm, (green triangles) 225 nm, (pink crosses) 412 nm. Bundle length is varied in an uncorrelated fashion by a factor of over two. Crosslinker axial spacing is calculated using a simple Langmuir isotherm approximation, $\delta = \delta_{min}(K_d + c_{fascin}) / c_{fascin}$ (45), where $\delta_{min} = 37.5$ nm is the minimum in-plane spacing between ABPs in hexagonally-ordered F-actin bundles (16) and $K_d = 0.5 \ \mu$M is the fascin-actin dissociation constant (46). Theoretical bundle

stiffness (solid curve) is calculated using Eq. (2) assuming $n = 27$, $\kappa_f = 4 \times 10^{-26}$ Nm$^2$, $c = 4$, $E_f A_f = 4.4 \times 10^{-8}$ N, and $k_\parallel \approx 10^{-5}$ N/m.

**Inset to 3B)** Dependence of the crossover values, $\alpha^{**}$, of the fiber coupling parameter on bundle size, $n$, at the decoupled-to-intermediate (bottom curves) and fully-coupled-to-intermediate (top curves) regime crossovers for (squares) pinned and (circles) clamped boundary conditions. Solid lines indicate $n$-independent and linear-in-$n$ scaling. Crossover values of $\alpha^{**}$ are defined by the value of $\alpha$ at which $\kappa_B$ is within a factor of two of its limiting decoupled and fully coupled values.

**Fig. 4** Bundle bending stiffness state-diagram for various biological and carbon-nanotube-based fiber bundles. Dashed red lines denote crossovers between (I) decoupled, (II) shear-dominated, and (III) fully coupled bending regimes. (a) Acrosomal process of the horseshoe crab sperm cell (11); (b) vertebrate hair cell stereocilia (1, 2, 35); (c) brush-border microvilli (1, 2, 47); (d) stress fibers; (e) filopodia (5); (f) drosophila neurosensory bristles (38); (g) outer pillar hair cell MT bundles (12); (h) uncrosslinked SWNT bundles (14); and (i) irradiation-crosslinked SWNT bundles (32). Spacing between ABPs is taken to be the minimal in-plane value for hexagonally-packed bundles, $\delta = 37.5$ nm (16). Extensional stiffnesses are, $L_f k_f = 4.4 \times 10^{-8}$ N, $2.6 \times 10^{-7}$ N, and $4.5 \times 10^{-7}$ N, for F-actin (21), MTs (20), and SWNTs (14), respectively.

**Fig. 1**

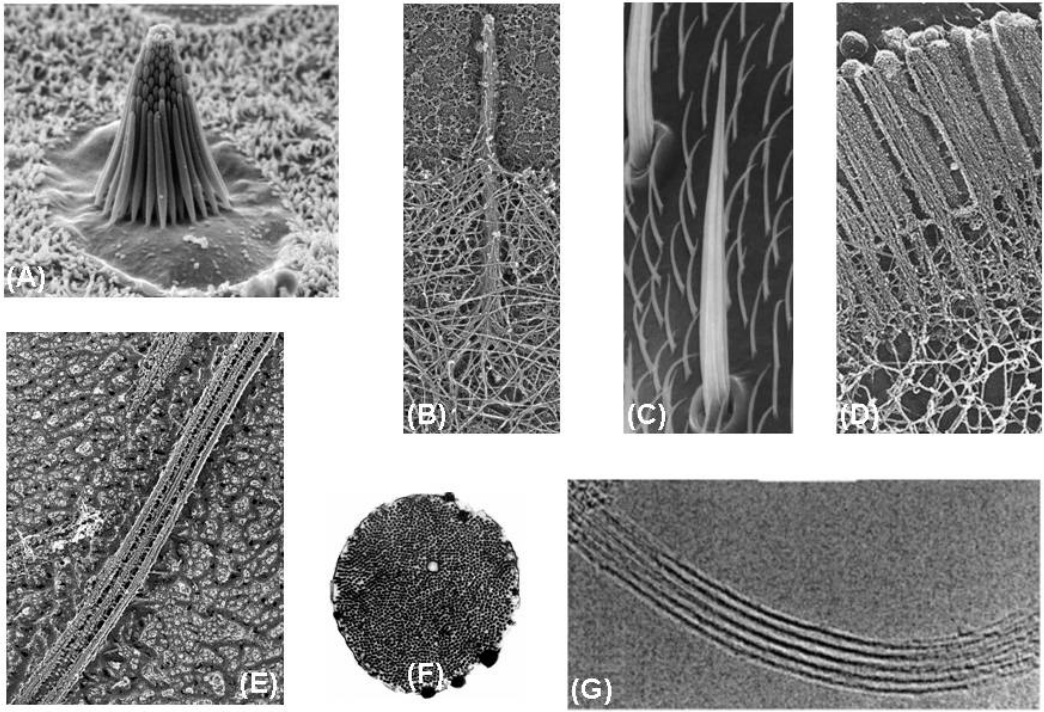

**Fig 2A**

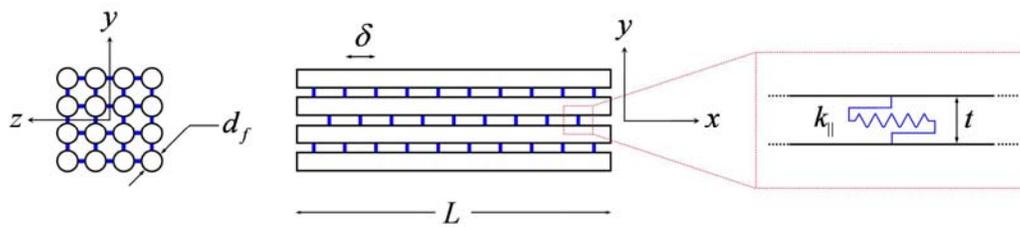

**Fig 2B**

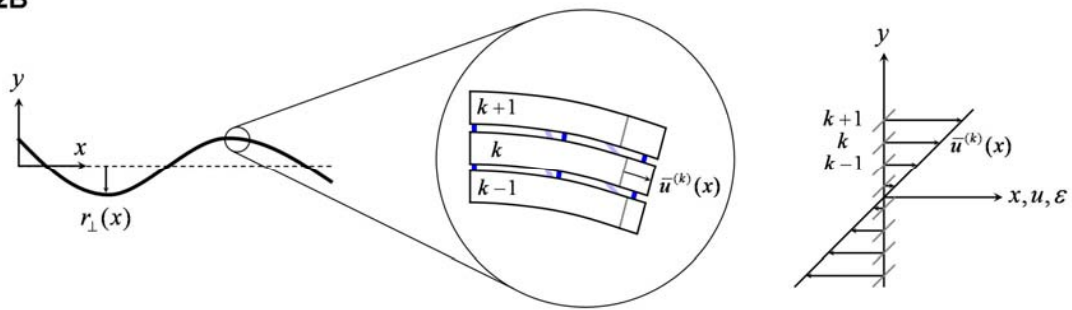

**Fig. 3A**          **Fig. 3B**          **Fig. 3C**

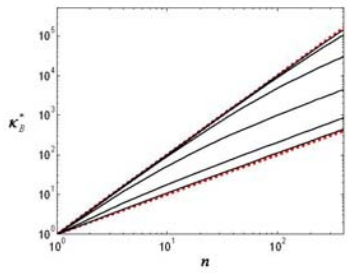 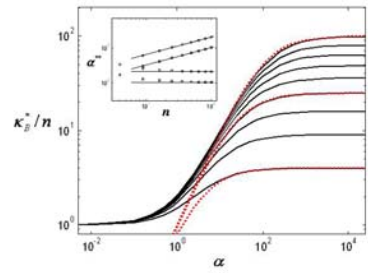 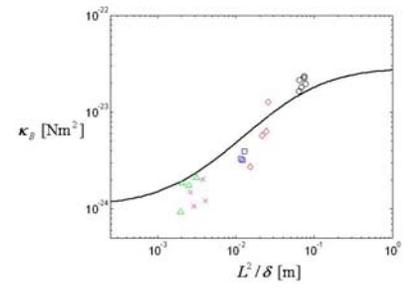

**Fig. 4**

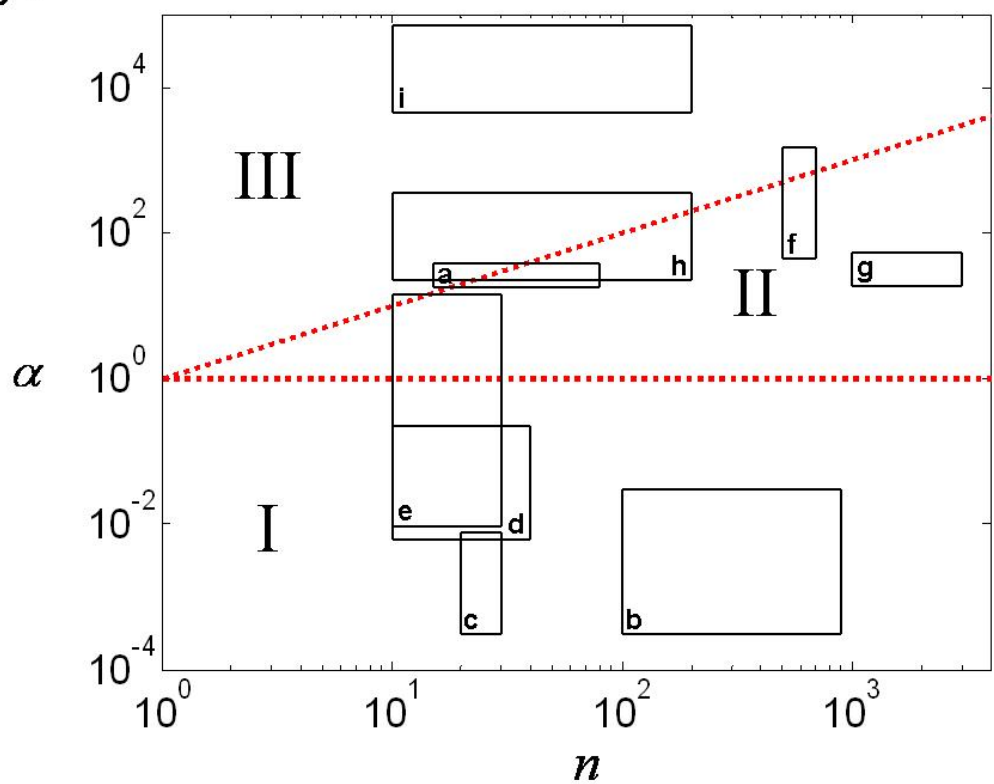